\documentclass[prl,twocolumn,showpacs,a4paper,superscriptaddress]{revtex4}
\usepackage{amsmath}
\usepackage{amsfonts}
\usepackage{epsfig}

\begin{document}
\title{Quantum Teleportation of Optical Quantum Gates} 
\author{Stephen D. Bartlett} 
\affiliation{Department of Physics, Macquarie University, Sydney, New
  South Wales 2109, Australia} 
\author{William J. Munro} 
\affiliation{Hewlett--Packard Laboratories, Filton Road, Stoke
  Gifford, Bristol BS34 8QZ, United Kingdom} 
\date{17 March 2003}

\begin{abstract}
  We show that a universal set of gates for quantum computation with
  optics can be quantum teleported through the use of EPR entangled
  states, homodyne detection, and linear optics and squeezing
  operations conditioned on measurement outcomes.  This scheme may be
  used for fault-tolerant quantum computation in any optical scheme
  (qubit or continuous variable).  The teleportation of
  nondeterministic nonlinear gates employed in linear optics quantum
  computation is discussed.
\end{abstract}
\maketitle

Quantum computation (QC) may allow certain problems to be solved more
efficiently than is possible on any classical machine~\cite{Nie00},
and optical realizations are particularly appealing because of the
advanced techniques of quantum optics for state preparation,
manipulation and measurement.  However, many challenges still remain
for fault-tolerant implementations of optical QC.  In proposals for QC
using optics~\cite{KLM01,Ral01,GKP01}, it is necessary to invoke some
form of an optical nonlinear transformation, and unlike linear optics
these nonlinear transformations either suffer from large losses
(decoherence) or employ nondeterministic gates that fail a large
fraction of the time.

The remarkable results by Gottesman and Chuang~\cite{Got99} show that
quantum teleportation~\cite{Ben93} can be used as a universal quantum
primitive.  In essense, quantum teleportation allows for the
fault-tolerant implementation of ``difficult'' quantum gates that
would otherwise corrupt the fragile information of a quantum state.
The linear optics quantum computation (LOQC) scheme~\cite{KLM01}
relies on the quantum teleportation of nondeterministic nonlinear
gates in the instances that they succeed in order to perform
fault-tolerant QC.  However, to overcome the difficulty in performing
Bell-state measurements in this optical encoding using only linear
optics~\cite{Lut99}, the LOQC scheme employs a near-deterministic
quantum teleportation that places severe demands on the photodetectors
employed for the measurements.  The requirements to perform
fault-tolerant quantum teleportation in this scheme greatly exceed
current technologies.

Fortunately, there exists an alternative approach for quantum
teleportation in optical systems: continuous-variable quantum
teleportation (CVQT)~\cite{Bra98,Fur98}.  The measurements involved in
this scheme are the highly developed and efficient techniques of
homodyne detection~\cite{Pol92}.  Also, experimental CVQT can be
performed unconditionally~\cite{Fur98}, and does not suffer from the
difficulties associated with the Bell-state measurements of qubit
quantum teleportation~\cite{Bra98d}.  It is of great interest, then,
to determine what optical quantum gates can be teleported using CVQT
(even in qubit schemes), and if such gates can be used to implement
fault-tolerant optical QC.

In this Letter, we show that CVQT can easily teleport quantum gates
that employ linear optics and squeezing operations.  Also, by allowing
squeezing operations conditioned on the results of measurements, we
show that one can teleport optical nonlinear gates generated by some
Hamiltonians that are cubic polynomial in the canonical coordinates.
We prove that the gates which can be teleported in this fashion form a
universal set of gates for QC, and that this scheme allows for the
fault-tolerant implementation of QC in \emph{any} optical scheme
employing qubits, qudits or CV encodings, provided one can perform
fault-tolerant linear optics and squeezing.  We analyze ``noisy''
teleportation employing finitely-squeezed states and imperfect
detectors, and the resulting effect on the teleported gates.  We
conclude by discussing the challenges involved in using CVQT to
teleport the nondeterministic gates in LOQC.

The transformations describing linear optics and squeezing possess a
natural group structure which will be exploited for their quantum
teleportation.  Thus, we begin by reviewing the Pauli and Clifford
groups for CV quantum computation~\cite{GKP01,Bar02}.  The standard
Pauli group $\mathcal{C}_1$ on $n$ coupled oscillator systems is the
Heisenberg-Weyl group, which consists of phase-space displacement
operators for the $n$ oscillators.  It is the group of ``linear
optics,'' which can be implemented by mixing states with strong
coherent fields at a beamsplitter~\cite{Wal94}.  The algebra that
generates this group is spanned by the $2n$ canonical operators
$\hat{q}_i$, $\hat{p}_i$, $i=1,\ldots,n$, along with the identity
operator $\hat{I}$, satisfying the commutation relations
$[\hat{q}_i,\hat{p}_j] = \text{i}\hbar \delta_{ij} \hat{I}$.  (In the
following, we set $\hbar=1$.)  For a single oscillator, the Pauli
group consists of operators of the form $R(q,p) = \exp(-\text{i}(q
\hat{p} - p \hat{q}))$, with $q,p \in \mathbb{R}$.  The Pauli
operators for one system can be used to construct a set of Pauli
operators $\{ R_i(q_i,p_i); i=1,\ldots,n \}$ for $n$ systems (where
each operator labeled by $i$ acts as the identity on all other systems
$j \neq i$).  This set generates the Pauli group $\mathcal{C}_1$.
The Clifford group $\mathcal{C}_2$ is the group of transformations
acting by conjugation that preserves the Pauli group $\mathcal{C}_1$;
i.e., a gate $U$ is in the Clifford group if $U R U^{-1} \in
\mathcal{C}_1$ for every $R \in \mathcal{C}_1$.  The Clifford group
$\mathcal{C}_2$ for continuous variables is easily shown~\cite{Bar02}
to be the (semidirect) product of the Pauli group and the linear
symplectic group of all one-mode and two-mode squeezing
transformations.

Transformations in the Clifford group do not form a universal set of
gates for optical QC.  However, Clifford group transformations
together with \emph{any} higher-order nonlinear transformation acting
on a single mode form a universal set of gates~\cite{Llo99}.  Because
arbitrary nonlinear gates can be constructed with such a set, this
result also applies to qubit-based optical realizations for any
encoding of qubits~\cite{KLM01,Ral01,GKP01}.  In the following, we
construct a scheme to use CVQT to teleport such a universal set of
gates.

\begin{figure}
  \includegraphics*[width=3.25in,keepaspectratio]{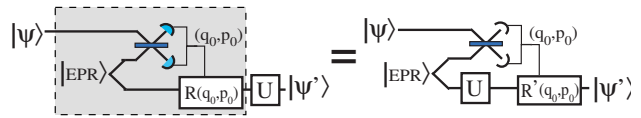}
  \caption{Quantum teleportation of a gate $U$.  The shaded box
    contains the CVQT circuit.  The process of first teleporting the
    state then implementing $U$ (resulting is the state $|\psi'\rangle
    = U|\psi\rangle$) is equivalent to acting on one mode of the EPR
    state with $U$, followed by a modified quantum teleportation.}
  \label{fig:TelGates}
\end{figure}

Quantum teleportation, although initially proposed for remote parties,
can also be used in a quantum circuit.  Consider a three mode optical
system (interferometer).  Let mode $1$ be in an arbitrary pure state
$|\psi\rangle$ (although this circuit works equivalently well for
mixed states), and modes $2$ and $3$ be in the maximally entangled
Einstein-Podolsky-Rosen (EPR) state $|{\rm EPR}\rangle_{23} = \int
|q\rangle_2 |q\rangle_3 \, {\rm d}q$, where $|q\rangle$ are position
eigenstates defined by $\hat{q}|q\rangle = q |q\rangle$.  Modes $1$
and $2$ are then subjected to joint projective measurements of the
form
\begin{equation}
  \label{eq:JointMeas}
  \Pi_{q,p} = R_1(q,p) |{\rm EPR}\rangle_{12}\langle{\rm EPR}|
  R^\dag_1(q,p) \, ,
\end{equation}
where $R_1(q,p)$ is the Pauli operator on mode~$1$.  The measurement
yields two classical numbers, $q_0$ and $p_0$, which are used to
condition a Pauli operation $R_3(q_0,p_0)$ on mode $3$.  The result of
this circuit is that mode $3$ is left in the state $|\psi\rangle$, and
we say that this state has been quantum teleported to mode $3$.  A
circuit diagram for quantum teleportation is given in the shaded box
of Fig.~\ref{fig:TelGates}.  Repeating this circuit on many modes
allows for the quantum teleportation of multi-mode states, preserving
entanglement between modes.

We now consider how CVQT can be used to teleport a quantum gate.
Consider an arbitrary transformation $S$ in the single mode ($n=1$)
Clifford group, and suppose we wish to implement this gate on an
arbitrary single mode state $|\psi\rangle$.  In the following, we show
that it is possible to implement $S$ on one mode of an EPR state, and
then employ a modified CVQT circuit to ``teleport'' this gate onto the
desired state $|\psi\rangle$.

We use the property that, if $U$ is in the single mode Clifford group,
then
\begin{displaymath}
  \label{eq:CommutingClifford}
  U R(q_0,p_0) = (U R(q_0,p_0) U^{-1}) U 
  = R'(q_0,p_0)U \, ,
\end{displaymath}
and $R'(q_0,p_0) = U R(q_0,p_0)U^{-1}$ is an element of the Pauli
group due to the definition of the Clifford group, determined by the
gate $U$.  Thus, the desired Clifford gate $U$ can be quantum
teleported onto the state $|\psi\rangle$ simply by implementing $U$ on
one mode of the EPR pair and by appropriately altering the conditional
displacement of the quantum teleportation.  Using $n$ single-mode
quantum teleportation circuits, it is then possible to teleport any
gate in the $n$-mode Clifford group.  See Fig.~\ref{fig:TelGates} for
an illustration of how this gate teleportation is performed.

In addition, this technique can be used to teleport more general
quantum gates than those in the Clifford group.  Consider the set of
gates defined by $\mathcal{C}_3 = \{ U | U\mathcal{C}_1 U^{-1}
\subseteq \mathcal{C}_2 \}$.  This set includes all gates that, when
conjugating any Pauli group operation, yield a Clifford group
operation.  Note that $\mathcal{C}_3$ is not a group (it is not closed
under composition) and contains nonlinear transformations.  One
example of a nonlinear gate in $\mathcal{C}_3$ is the \emph{cubic
  phase gate} $V(\gamma) = \exp (\text{i} \gamma \hat{q}^3 )$, with
$\gamma \in \mathbb{R}$, which Gottesman \emph{et al}~\cite{GKP01}
have shown how to implement using Clifford group transformations and
homodyne and photon counting measurements.  (For details,
see~\cite{Bar02b}.) An example of a multimode gate in $\mathcal{C}_3$
is the \emph{controlled phase gate} $CP_{ij} =
\exp(\frac{1}{2}\text{i} \hat{q}_i \hat{q}_j^2)$ acting on two modes
$i$ and $j$.  In comparison to the CV phase gate~\cite{Bar02}, this
gate performs a phase gate operation on mode $j$, the magnitude of
which depends on the state of mode $i$.  It is a controlled Clifford
group transformation; others can be defined similarly.

We now define a quantum teleportation circuit to teleport a gate $U
\in \mathcal{C}_3$.  Commuting this gate through the conditional
operations gives
\begin{displaymath}
  \label{eq:CommutingC3}
  U R(q_0,p_0) = (U R(q_0,p_0) U^{-1}) U = R_2'(q_0,p_0) U \, ,
\end{displaymath}
where, by the definition of $\mathcal{C}_3$, $R_2'(q_0,p_0)$ is an
element of the Clifford group.  Thus, any gate in $\mathcal{C}_3$ can
be quantum teleported using Clifford group transformations conditioned
on measurement outcomes; i.e., conditional phase-space displacements
and squeezing operations.

Consider, as an example, the CVQT of the cubic phase gate $V(\gamma)$.
Commuting this gate back through the Pauli operator gives
\begin{align*}
  \label{eq:NewConditionalVgamma}
  R_2'(q_0,p_0;\gamma) &= V(\gamma) R(q_0,p_0) V(\gamma)^{-1} \\
  &= \exp\bigl(-\text{i} \bigl(q_0
  \hat{p} - p_0 \hat{q} + 3 \gamma q_0 \hat{q}^2 \bigr) \bigr) \, .
\end{align*}
This operation, generated by an inhomogeneous quadratic polynomial in
$\hat{q}$ and $\hat{p}$, lies in the Clifford group and can be
implemented using a combination of conditional phase-space
displacements and one-mode squeezing.  Thus, to teleport $V(\gamma)$,
this gate is performed on one mode of an EPR state, followed by a
modified teleportation scheme with a conditional operation
$R_2'(q_0,p_0;\gamma)$; the result is that a state $|\psi\rangle$ is
quantum teleported into the transformed state $V(\gamma)|\psi\rangle$.

The cubic phase gate $V(\gamma)$, being a higher-order nonlinear gate
on a single mode, can be combined with Clifford group gates of $n$
modes to form a universal set of gates for QC on $n$
modes~\cite{Llo99}, and a scheme exists to implement this cubic phase
gate~\cite{GKP01}.  Thus, with CVQT it is possible to teleport a
universal and realizable set of gates.  If Clifford group
transformations and CVQT can be implemented fault-tolerantly, then it
is possible to use this scheme to implement a fault-tolerant cubic
phase gate (or other nonlinear gate in $\mathcal{C}_3$) using a gate
that is not fault-tolerant.  Thus, any nonlinear transformations can
be moved ``off-line''~\cite{GKP01,Got99}; although these
transformations must still be performed in the quantum teleportation
circuit, they can be made to act on EPR ancilla states
non-deterministically rather than on the fragile encoded states.

Because most optical quantum information schemes employ the Kerr
effect (generated by a Hamiltonian of the form $(\hat{a}^\dag)^2
\hat{a}^2$) as the nonlinear transformation outside of the Clifford
group, it is of interest to consider how such a transformation can be
implemented using the above universal set of gates.  Using the
relation $e^{\text{i}At}e^{\text{i}Bt}e^{-\text{i}At}e^{-\text{i}Bt} =
e^{\text{i}[A,B]t^2} + \mathcal{O}(t^3)$, a combination of cubic phase
gates and Clifford gates can be used to simulate the Kerr nonlinearity
to any degree of accuracy.  We also note that it is possible to
iteratively define more higher-order sets of gates $\mathcal{C}_k =
\{U|U\mathcal{C}_1 U^{-1} \subseteq \mathcal{C}_{k-1} \}$ that can be
implemented fault-tolerantly if gates in $\mathcal{C}_{k-1}$ can be
implemented fault-tolerantly.  The Kerr nonlinearity is related to
transformations in the set $\mathcal{C}_4$.

We now consider the effects of realistic noise and errors that may
occur in this scheme, and show that quantum error correcting codes
(QECCs) may allow for fault-tolerant quantum computation.  Clifford
group transformations describing linear optics and squeezing will in
general contain errors such as imprecise transformations (e.g.,
imperfect beamsplitter reflectivities) or linear coupling to the
environment such as amplitude damping.  Due to the linear nature of
these couplings, such errors are described by ``small'' Clifford group
transformations; it is unlikely that a large nonlinear error will
occur in an optical system.  For example, errors in the implementation
of a displacement operation may be described by small displacement
errors $R(\delta q,\delta p)$ with, say, a Gaussian distribution; see
below.  Thus, we represent such errors in Clifford group
transformations by distributions of linear transformations
$R(\delta q,\delta p)$, where $\delta q^2 + \delta
p^2 < S^2$ for some maximum error distance $S$ in phase space.

To describe errors in CVQT, we must take into account finite squeezing
and imperfect detection.  In experiment~\cite{Fur98}, the EPR states
are approximated by two-mode squeezed vacua~\cite{Enk99}
\begin{equation}
  \label{eq:SqueezedVac}
  |\eta\rangle_{23} = \sqrt{1-\eta^2}
  \sum_{n=0}^\infty \eta^n |n\rangle_2 |n\rangle_3 \, ,
\end{equation}
with $|n\rangle$ the $n$-boson Fock state and $0 < \eta \leq 1$.  In
the limit $\eta \to 1$, this state becomes the maximally entangled EPR
state $|{\rm EPR}\rangle_{23}$.  The projective measurements of
Eq.~(\ref{eq:JointMeas}) are implemented via homodyne detection, and
the displacement operation $R_3(q_0,p_0)$ is performed by mixing the
field with a local oscillator at a beamsplitter;
see~\cite{Bra98,Fur98} for details.  Finite squeezing $\eta < 1$ and
imperfect homodyne efficiency $\nu < 1$ can characterised by a
Gaussian noise term with variance $\sigma = \exp(-2\tanh^{-1}\eta) +
(1-\nu^2)/\nu^2$, defined such that $\sigma = 1/2$ is the level of
vacuum noise~\cite{Joh02}.  For demonstrated CVQT with fidelity $F >
0.5$, we require $\sigma<1$; for fault-tolerant teleportation of
quantum gates, $\sigma$ must be significantly smaller unless
appropriate quantum error correction can be applied (see below).

Imperfect quantum teleportation can be described by a \emph{transfer
  superoperator} $\mathcal{E}_\sigma$ defined on a state $\rho$ as
\begin{displaymath}
  \mathcal{E}_\sigma (\rho) = \int \frac{{\rm d} q\, {\rm d}
  p}{\pi\sigma} \exp(-\frac{q^2 + p^2}{\sigma}) R(q,p)\rho R^\dag(q,p)\, .
\end{displaymath}
The transfer superoperator has the effect of convoluting the state
$\rho$ with a Gaussian of standard deviation $\sigma$.  Let
$\mathcal{E}_U$ be the superoperator corresponding to a unitary
transformation $U$, defined on a density matrix $\rho$ to be
$\mathcal{E}_U(\rho) = U \rho U^{-1}$.  Using noisy quantum
teleportation to teleport the gate $U$, the resulting transformation
is described by the superoperator $\mathcal{E}_U \circ
\mathcal{E}_\sigma$.  If $\mathcal{E}'_U$ describes a gate
implementing the unitary transformation $U$ with small linear noise,
the superoperator $\mathcal{E}'_U \circ \mathcal{E}_\sigma$ describes
a teleported gate with an additional Gaussian noise due to the
teleportation.

Thus, noisy quantum teleportation of gates is described by a
convolution of the (possibly noisy) gate with a Gaussian noise
process; the corresponding errors can be viewed as random
displacements $R(\delta q, \delta p)$ with $\delta q^2 + \delta p^2
\sim \sigma$.  The effect of this Gaussian noise and the existence of
QECCs to correct this noise depends critically on the specific
encoding used.  Single photon encodings~\cite{KLM01} and
superpositions of coherent states~\cite{Ral01} are extremely fragile
to Gaussian noise.  The encoding of Gottesman \emph{et
  al}~\cite{GKP01}, however, is specifically designed to protect
against such Gaussian noise because small displacements in phase space
can be detected and corrected.  Briefly, this encoding employs
superpositions of highly squeezed states, for which small displacement
errors $R(\delta q, \delta p)$ can be detected and corrected using
only Clifford group transformations and homodyne detection.  As long
as the displacement errors in the Clifford group transformations and
the CVQT are sufficiently small compared to the length scale (in phase
space) of the encoding, quantum error correction can be applied
frequently enough to keep these errors small and contained;
see~\cite{GKP01} for details.  In addition, these codes can be
concatenated with the QECCs of~\cite{Bra98b}, which protect against
large displacement errors on a single mode.  Thus, our proposal for
the quantum teleportation of nonlinear gates can be made
fault-tolerant if encodings are used that are protected against
Gaussian noise and the QECCs use fault-tolerant Clifford group
operations (i.e., operations with small errors in the above sense).
Creating the encoded states of~\cite{GKP01} is extremely challenging
because it requires coherent superpositions of highly squeezed states.
These Gaussian noise processes, however, are indicative of optical
systems, and the investigation of ``Gaussian protected'' encodings
that require only Clifford group operations is essential for robust
quantum information processing using optics.

Because CVQT can teleport \emph{any} state of an optical mode, we can
consider the use of CVQT in any optical quantum information process,
even qubit-based schemes.  One scheme~\cite{Ral01} encoding qubits as
coherent states could employ CVQT to implement the ``difficult''
Hadamard transformations requiring a Kerr nonlinearity in a
fault-tolerant way.  In LOQC, nondeterministic gates are used to
implement nonlinear transformations on arbitrary states with up to two
photons.  Qubit-based quantum teleportation is employed to implement
these gates fault-tolerantly, but the near-deterministic quantum
teleportation of LOQC places strong demands on photodetectors and
requires highly entangled multimode states.  To instead employ CVQT,
the problem arises that any gate to be teleported must be made to act
on one mode of an EPR state.  The probability of a successful gate
operation may be affected by the photon number cutoff used for these
states (the nonlinear sign gate being one such example).  For
high-fidelity quantum teleportation, one must employ highly squeezed
states that have a corresponding high photon number cutoff (see
Eq.~(\ref{eq:SqueezedVac})); thus, any effective teleportable gate
must be designed to operate on such high-photon states.  However, the
challenges to design such a non-deterministic gate that can be
teleported using this scheme may be less significant than the
difficulties involved in the near-deterministic teleportation of LOQC.

In conclusion, we have shown that fault-tolerant implementations of
Clifford group transformations and CVQT is sufficient to perform
fault-tolerant QC in any optical scheme: qubits, qudits or CV.  A
nonlinear transformation such as the cubic phase gate (possible using
Clifford group transformations, homodyne detection, and photon
counting) can be performed ``off-line'' on EPR states and used as a
resource to perform universal QC.  Such gates would allow the
implementation of difficult but critical nonlinear transformations
such as the Kerr effect in a fault-tolerant way.  The definition of an
infinite series of gates $\mathcal{C}_k$ highlights which gates can be
teleported in a straightforward manner.  In particular, the cubic
phase gate (rather than a Kerr nonlinearity) is identified as the most
direct nonlinear operation that would achieve a universal set of
gates.  The simplicity and power of this gate clearly motivates its
experimental demonstration and its use in future quantum information
processing tasks.  The use of CVQT in LOQC relies on the design of new
nondeterministic nonlinear gates that can be teleported using this
scheme.  Finally, the Gaussian noise introduced through realistic CVQT
can be corrected with suitable ``Gaussian protected'' encodings and
Clifford group transformations, and the results presented here
highlight the need for new, realistic quantum error correction schemes
of this form.

\begin{acknowledgments}
  This work was supported in part by the European Union through RAMBOQ
  and the QUIPROCONE Network, and by Macquarie University.  We
  acknowledge helpful discussions with K.\ Nemoto, T.\ Ralph and B.\ 
  C.\ Sanders.
\end{acknowledgments}

\end{document}